\documentclass[aps,prl,groupedaddress,twocolumn,superscriptaddress]{revtex4-1}
\usepackage{graphicx}
\usepackage{amsmath}
\usepackage{color,soul,array}

\newcommand{\TN}{$T_{\mathrm{N}}$}
\newcommand{\Gfour}{$\Gamma^{\mathrm{4g}}$}
\newcommand{\Gfive}{$\Gamma^{\mathrm{5g}}$}

\begin{document}


\title{Multi-site exchange enhanced barocaloric response in Mn$_{3}$NiN}


\author{David Boldrin}
\email[Corresponding author: d.boldrin@imperial.ac.uk]{}
\affiliation{Department of Physics, Blackett Laboratory, Imperial College London, London, SW7 2AZ, UK}
\author{Eduardo Mendive-Tapia}
\affiliation{Department of Physics, University of Warwick, Coventry CV4 7AL, United Kingdom}
\author{Jan Zemen}
\affiliation{Faculty of Electrical Engineering, Czech Technical University in Prague, Technická 2, Prague 166 27, Czech Republic}
\author{Julie B. Staunton}
\affiliation{Department of Physics, University of Warwick, Coventry CV4 7AL, United Kingdom}
\author{Thomas Hansen}
\affiliation{Institut Laue-Langevin, CS 20156, 38042 Grenoble, Cedex 9, France}
\author{Araceli Aznar}
\affiliation{Departament de F\'{i}sica, EEBE, Campus Diagonal-Bes\`{o}s and Barcelona Research Center in Multiscale Science and Engineering, Universitat Politècnica de Catalunya, Eduard Maristany, 10-14, 08019 Barcelona, Catalonia, Spain}
\author{Josep-Llu\'{i}s Tamarit}
\affiliation{Departament de F\'{i}sica, EEBE, Campus Diagonal-Bes\`{o}s and Barcelona Research Center in Multiscale Science and Engineering, Universitat Politècnica de Catalunya, Eduard Maristany, 10-14, 08019 Barcelona, Catalonia, Spain}
\author{Maria Barrio}
\affiliation{Departament de F\'{i}sica, EEBE, Campus Diagonal-Bes\`{o}s and Barcelona Research Center in Multiscale Science and Engineering, Universitat Politècnica de Catalunya, Eduard Maristany, 10-14, 08019 Barcelona, Catalonia, Spain}
\author{Pol Lloveras}
\affiliation{Departament de F\'{i}sica, EEBE, Campus Diagonal-Bes\`{o}s and Barcelona Research Center in Multiscale Science and Engineering, Universitat Politècnica de Catalunya, Eduard Maristany, 10-14, 08019 Barcelona, Catalonia, Spain}
\author{Jiyeob Kim}
\affiliation{Department of Materials Science, University of Cambridge, Cambridge CB3 0FS, United Kingdom}
\author{Xavier Moya}
\affiliation{Department of Materials Science, University of Cambridge, Cambridge CB3 0FS, United Kingdom}
\author{Lesley F. Cohen}
\affiliation{Department of Physics, Blackett Laboratory, Imperial College London, London, SW7 2AZ, UK}

\date{\today}

\begin{abstract}
We have studied the barocaloric effect (BCE) in the geometrically frustrated antiferromagnet Mn$_{3}$NiN across the N\'{e}el transition temperature. Experimentally we find a larger barocaloric entropy change by a factor of 1.6 than that recently discovered in the isostructural antiperovskite Mn$_{3}$GaN despite greater magnetovolume coupling in the latter. By fitting experimental data to theory we show that the larger BCE of Mn$_{3}$NiN originates from multi-site exchange interactions amongst the local Mn magnetic moments and their coupling with itinerant electron spins. Using this framework, we discuss the route to maximise the BCE in the wider Mn$_{3}$AN family. 
\end{abstract}

\maketitle

\section{Introduction}

The emerging field of solid-state caloric cooling offers opportunities for greater energy-efficient refrigeration without the need for environmentally harmful chemicals.  Magnetocaloric properties are by far the most studied, whereas mechanocalorics, including baro- and elastocalorics are only beginning to gain prominence \cite{Manosa2017}. In the former, the largest effects are found in materials with contributions from both localized and itinerant magnetic moments so-called mixed magnetism with competing exchange interactions. In the intermetallics $R$Co$_{2}$ ($R$ = rare-earth) \cite{Singh2007}, Mn-Fe(P,Si) \cite{Dung2011} and La(Fe,Si)$_{13}$ \cite{Fujita1999,Gruner2015}, mixing of magnetic contributions and a strong coupling to the crystal lattice results in 1$^{\mathrm{st}}$-order transition behaviour, whilst maintaining large magnetisation and a high transition temperature. These properties also lend themselves to attractive barocaloric effects (BCE) and this complimentary behaviour has guided initial research in this field \cite{Moya2014,Yuce2012,Manosa2010,Manosa2011,Stern-Taulats2014}. Recent efforts have also focused on qualitatively different material families such as ferrielectric inorganic salts \cite{Lloveras2015} and hybrid inorganic-organic materials \cite{Bermudez-Garcia2017}. Whilst impressive BCE are achieved in these materials, low density of the materials, poor thermal conductivity and long-term stability are issues that may limit their applicability. An alternative and promising family of magnetic systems which we explore here are geometrically frustrated antiferromagnets (AFM) \cite{Matsunami2014}. The most prominent example is the metallic alloy Mn$_{3}$GaN where the BCE effect is thought to be enhanced due to the stabilization of the local moment in the non-collinear AFM phase \cite{Matsunami2014}. Thus, the combination of geometric frustration of exchange between local Mn moments and itinerant electrons, akin to the local and itinerant character of the optimum magnetocalorics, may provide a fruitful playground through which to explore new barocaloric materials.

The Mn$_{3}$AN family displays a number of unusual properties, such as anomalous coefficient of resistivity \cite{Chi2001}, negative thermal expansion \cite{Deng2015} and piezomagnetism \cite{Zemen2017,Zemen2017a,Boldrin2018}. The origin of these properties lies in a combination of electronic features: (i) large local magnetic moments due to a half-filled $d$-state, (ii) a non-collinear AFM order born from frustrated Mn-Mn magnetic interactions and (iii) the mixing of itinerant and localised spin degrees of freedom and the self-sustaining interactions that occur between them \cite{Matsunami2014,Zemen2017,Staunton2014}. The two non-collinear magnetic structures commonly found in the Mn$_{3}$AN family are shown in Figs. \ref{zero-pressure-fig}(a) and \ref{zero-pressure-fig}(b) and will be hereafter termed \Gfour\ and \Gfive, respectively. 
These structures are similar to that which underlies the large anomalous Hall effect in the AFM Mn$_{3}$Sn, the only difference being the sign of the chirality \cite{Nakatsuji2015}.
The intrinsic frustration that leads to these non-collinear structures, combined with the energetic stability of their cubic lattice with respect to structural perturbations of lower symmetry, results in large relative volume changes, $\Delta\omega_{T_{\mathrm{N}}}$, at the 1$^{\mathrm{st}}$-order N\'{e}el transition temperature, \TN, and a relative insensitivity of the latter to pressure, $p$. From these properties, a BCE figure-of-merit (FOM), $\Delta\omega_{T_{\mathrm{N}}}$\,$|\frac{\mathrm{d}T_{\mathrm{N}}}{\mathrm{d}p}|^{-1}$ \cite{Matsunami2014}, has been proposed that allows comparisons to be made between material families. Here we measure the BCE in the closely related Mn$_{3}$NiN. Despite its smaller magnetovolume coupling compared to Mn$_{3}$GaN, we find larger barocaloric entropy changes under hydrostatic  pressure. We find the enhanced barocaloric properties in Mn$_{3}$NiN can be explained in terms of multi-site exchange rather than larger magnetovolume coupling. This insight has implications for improved barocaloric materials going forward.

\section{Ambient Pressure Measurements}

We will first discuss measurements performed on Mn$_{3}$NiN at ambient pressure. The sample was prepared using a standard solid-state synthesis technique, as described previously \cite{Boldrin2017}. Neutron powder diffraction data were collected on the D20 diffractometer ($\lambda = 1.544$\,\AA\ ) at ILL, France \cite{BoldrinILL}. Rietveld refinement of the data confirmed that the sample was the antiperovskite Mn$_{3}$NiN phase with only a minor MnO impurity (Fig. S1). The refined lattice parameter at room temperature was 3.88075(3)\,\AA, in excellent agreement with the literature \cite{Fruchart1971}, and all atomic sites were refined to within 98\,\% occupation, thus confirming that the sample was close to stoichiometric with a fully occupied N site. The latter is particularly noteworthy as nitrogen deficiency is reported to occur in manganese nitride compounds \cite{Takenaka2014}.

\begin{figure}
\includegraphics[width=0.45\textwidth]{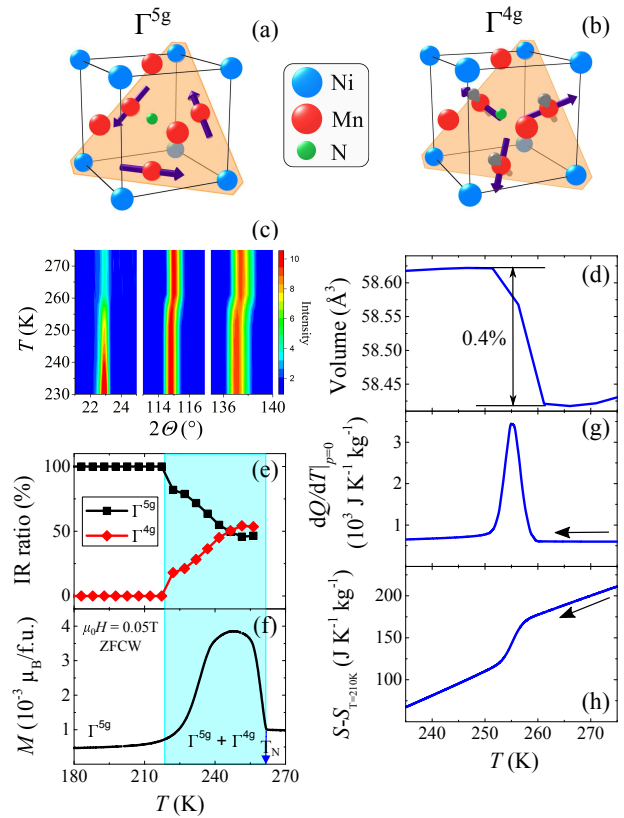}%
\caption{(a) \Gfive\ and (b) \Gfour\ magnetic structures of Mn$_{3}$NiN. \Gfive\ is a fully compensated antiferromagnetic structure stable at low temperatures. \Gfour\ is also an antiferromagnetic structure (purple arrows), but has a symmetry allowed ferromagnetic component (grey arrows) causing canting of the spins out of the (111) plane. (c) Thermodiffractogram measured using neutron powder diffraction ($\lambda = 1.544$\,\AA\ ). (d) Unit cell volume as a function of temperature refined from the neutron diffraction data. (e) The refined contributions of the \Gfive\ and \Gfour\ representations to the magnetic structure. (f) Bulk magnetometry data collected on the same sample under an applied field $\mu_{0}H = 0.05$\,T after zero field cooling. The blue highlighted region indicates the enlarged moment associated with the ferromagnetic contribution allowed in the \Gfour\ representation. (g) Temperature dependent heat flow at ambient pressure, $\frac{\mathrm{d}Q}{\mathrm{d}T}|_{p=0}$, which is equal to heat capacity outside the transition region, and includes contributions that arise from latent heat within, and (h) total entropy change calculated from (g) for the same sample on cooling.}
\label{zero-pressure-fig}
\end{figure}

The temperature dependence of the neutron diffraction data is summarised in Fig. \ref{zero-pressure-fig}(c). At \TN, a clear shift in the nuclear Bragg peaks indicates a significant change in lattice parameters concomitant with the appearance of magnetic Bragg peaks. The refined lattice parameter across the transition reveals a volume change of $\Delta\omega_{T_{\mathrm{N}}} = 0.4$\,\% (Fig. \ref{zero-pressure-fig}(d)), which is similar to that found in previous studies on this material \cite{Takenaka2014,Wu2013}. The magnetic Bragg peaks are centred on the nuclear peaks, thus confirming a $k = 0$ magnetic propagation vector. The temperature dependence of the magnetic structure is similar to that found previously \cite{Fruchart1971}. At \TN\ the magnetic moments are rotated within the (111) plane roughly between the \Gfour\ and \Gfive\ structures and upon further cooling they rotate towards the \Gfive\ structure until $\sim 220$\,K when this rotation is complete (Fig. \ref{zero-pressure-fig}(e)). This temperature window coincides with an enlarged magnetisation evidenced from the bulk magnetometry data (Fig. \ref{zero-pressure-fig}(f)), which can be explained by the symmetry allowed ferromagnetic component of the \Gfour\ structure (Fig. \ref{zero-pressure-fig}(b)) \cite{Fruchart1971}.

Evidence of the 1$^{\mathrm{st}}$-order character of the transition is found in the latent heat ($Q_{\mathrm{t}} \sim T_{\mathrm{N}}\Delta S_{\mathrm{t}}$) from the heat flow data shown in Fig \ref{zero-pressure-fig}(g). The total entropy change in zero pressure, $S-S_{210\,\mathrm{K}}$, calculated from the same data as $S-S_{210\,\mathrm{K}} = \int^{T}_{210\,\mathrm{K}} \left(1/T(\frac{\mathrm{d}Q}{\mathrm{d}T})_{p=0}\right)$\,$\mathrm{d}T$, is presented in Fig. \ref{zero-pressure-fig}(h). From the latter we find the entropy change across \TN, $|\Delta S_{\mathrm{t}}|$, at zero pressure is 43\,J\,K$^{-1}$\,kg$^{-1}$, approximately double the value found in Mn$_{3}$GaN \cite{Matsunami2014}.

\begin{figure}
\includegraphics[width=0.4\textwidth]{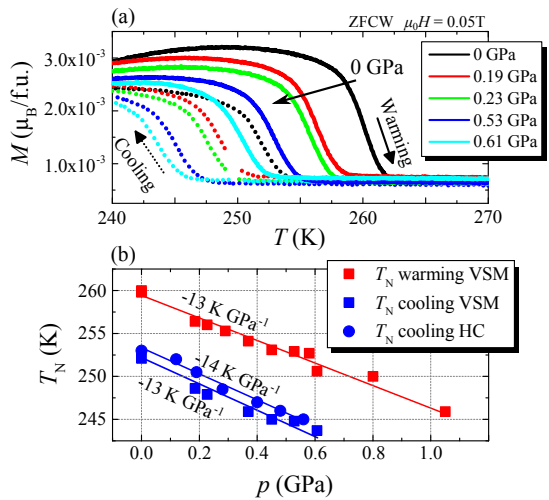}%
\caption{(a) Magnetisation under an applied field of $\mu_{0}H = 0.05$\,T, as a function of temperature and measured at different pressures after zero field cooling. (b) Pressure dependence of \TN, taken from the derivative of the $M(T)$ VSM data and the peak of the HC data. Solid lines are linear fits to the data.}
\label{pressure-fig}
\end{figure}

\section{Pressure Dependent and Barocaloric Effect Measurements}

We now  turn to measurements performed under pressure (experimental details in S.I.). The vibrating sample magnetometry (VSM) data of Fig. \ref{pressure-fig}(a) shows that $T_{\mathrm{N}}$ shifts to lower $T$ with increased pressure. Moreover, in zero pressure the thermal hysteresis is 8\,K whilst under a pressure of 0.61\,GPa the hysteresis is reduced to 7\,K and the transition is broadened noticeably. As shown in Fig. \ref{pressure-fig}(b), the pressure sensitivity of \TN\ ranges between $-13\pm1$ and $-14\pm1$\,K\,GPa$^{-1}$, as determined by either high-pressure calorimetry (HC, explained below) or VSM, showing that they are in excellent agreement.
  
The results under pressure allow us to make a direct comparison with available data on the Mn$_{3}$GaN antiperovskite \cite{Matsunami2014}. Firstly, $\Delta\omega_{T_{\mathrm{N}}}$ at the transition of 0.4\,\% is roughly half the value of that in the Ga system \cite{Takenaka2014}. However, the sensitivity of \TN\ with pressure, $|\frac{\mathrm{d}T_{\mathrm{N}}}{\mathrm{d}p}|$, of $13.5$\,K\,GPa$^{-1}$ (taken from the average of the measured values) is $\sim 5$ times smaller. If the magnetovolume coupling were the primary underlying mechanism for 1$^{\mathrm{st}}$-order character, one would expect $\Delta\omega_{T_{\mathrm{N}}}$ and $|\frac{\mathrm{d}T_{\mathrm{N}}}{\mathrm{d}p}|$ to scale linearly, given the chemical and magnetic similarities of the two systems and that both parameters are heavily dependent on the strength of the magnetovolume coupling. By combining these values we find the FOM, $\Delta\omega_{T_{\mathrm{N}}} |\frac{\mathrm{d}T_{\mathrm{N}}}{\mathrm{d}p}|^{-1}$ \cite{Matsunami2014}, in Mn$_{3}$NiN is 0.03\,\%\,GPa\,K$^{-1}$, more than double that of Mn$_{3}$GaN and several other known barocaloric systems \cite{Matsunami2014}.

\begin{figure}
\includegraphics[width=0.45\textwidth]{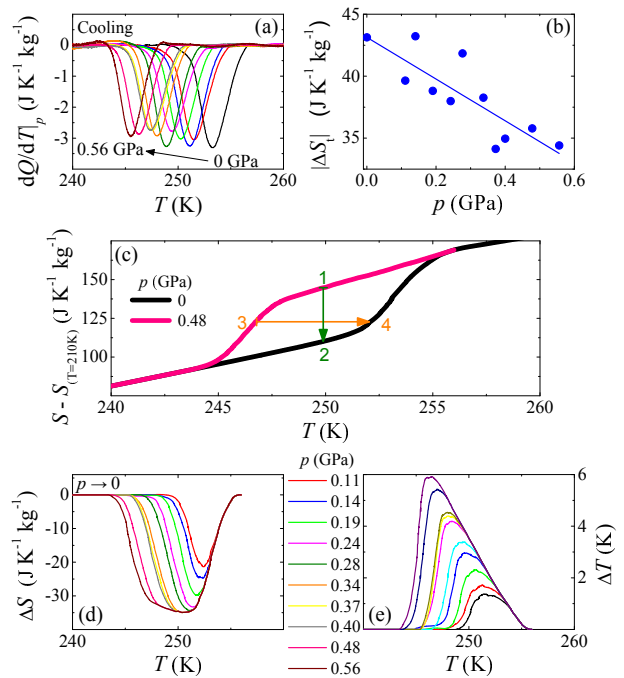}%
\caption{\label{}(a) Temperature dependent heat flow under pressure, $\frac{\mathrm{d}Q}{\mathrm{d}T}|_{p}$, on cooling through the transition at different values of increasing pressure, $p$, after baseline subtraction. The full legend is the same as in (d) and (e). (b) Entropy changes cooling through \TN\ obtained from the data in (a). (c) Isobaric entropy curves, $S-S_{210\,\mathrm{K}}$, on cooling at 0 and 0.48\,GPa applied pressure. The green arrow from 1 to 2 and orange arrow from 3 to 4 indicate the isothermal entropy and adiabatic temperature changes, respectively, upon removal of pressure. (d-e) Isothermal entropy and adiabatic temperature changes obtained from data in (a) as explained in (c).}
\label{BCE-fig}
\end{figure}

We determine the BCE by quasi-direct caloric measurements under pressure (experimental details in S.I.). Temperature dependent heat flow measurements under pressure, $\frac{\mathrm{d}Q}{\mathrm{d}T}|_{p}$, recorded on cooling (Fig. \ref{BCE-fig}(a)) show a sharp peak at \TN\ that shifts to lower temperatures with applied pressure. Figure \ref{BCE-fig}(b) shows that the entropy change under applied pressures decreases, falling to a value upon cooling of $34$\,J\,K$^{-1}$\,kg$^{-1}$ in 0.56\,GPa. This decrease is not due to additional entropy changes associated with the thermal expansion on either side of the transition \cite{Lloveras2015}, as these are negligible (Fig. \ref{zero-pressure-fig}(d)). One explanation could be that the transition broadens (becomes weakly 1$^{\mathrm{st}}$-order) with increased pressure, which is consistent with the magnetometry data (Fig. \ref{pressure-fig}(a)), although a detailed understanding would require further study.

To calculate pressure-driven isothermal changes in entropy, $\Delta S$, and pressure-driven adiabatic changes in temperature, $\Delta T$, we use the total entropy curves, $S-S_{210\,\mathrm{K}}$ (Fig. \ref{BCE-fig}(c)), measured under various applied pressures. The isothermal entropy changes, Fig. \ref{BCE-fig}(d), are calculated upon the release of pressure to atmospheric. The peak value of $|\Delta S|$ increases to a value of $\sim 35$\,J\,K$^{-1}$\,kg$^{-1}$ by removing 0.28\,GPa, and subsequently saturates upon removing pressure from higher values, whereas the temperature window at which the values of $|\Delta S|$ are large increases continuously with removal of higher pressures. This maximum value of $|\Delta S|$ is $\sim 1.6$ times larger than Mn$_{3}$GaN, in approximate agreement with the relative FOM values.  On the other hand, for $\Delta T$ both the peak value and temperature range for which $\Delta T > 0$ continuously increase upon removal of the largest applied pressure of 0.56\,GPa. The largest peak value of $\Delta T$ is 5.8\,K, therefore the available adiabatic temperature change for a given pressure release, $\frac{\Delta T}{\mathrm{d}p}$, is 10.4\,K\,GPa$^{-1}$. Thus, despite the enhanced entropy changes in Mn$_{3}$NiN the $\Delta T$ extracted from experiment appears modest. Note that in order to drive these BCE in a reversible manner, pressures larger than 0.56\,GPa would be required.

\section{Theoretical Modelling}

The lack of scaling between $\Delta\omega_{T_{\mathrm{N}}}$ and $|\frac{\mathrm{d}T_{\mathrm{N}}}{\mathrm{d}p}|$ suggests properties beyond magnetovolume coupling contribute to $|\Delta S|$. To further understand the giant barocaloric entropy changes in Mn$_{3}$NiN we carried out an analysis of our experimental data based on a description of the Gibbs free energy, $G$.
We designed a simple generic form of $G$ drawing from our previous, detailed ab-initio DFT-DLM theory modelling of the complex magnetic phase diagrams of the heavy rare earth metals \cite{Mendive-Tapia2017}, transition metal alloys such as FeRh \cite{Staunton2014} and pertinent to this work Mn$_{3}$GaN \cite{Zemen2017}. This form of $G$ is produced by using the Feynman-Peierls' Inequality~\cite{Feynman1955} and a trial 'spin' Hamiltonian. 
\begin{equation}
\mathcal{H}_0(\{\hat{e}_n\})
=-\sum_n{\textbf{h}_n\cdot\hat{e}_n},
\label{EQ1-Theo}
\end{equation}
 We solve the statistical mechanics of Eq.\ (\ref{EQ1-Theo}) \cite{Gyorffy1985} by finding the probability
\begin{equation}
P_0(\{\hat{e}_n\})=\frac{1}{Z_0}\exp\left[-\beta\mathcal{H}_0(\{\hat{e}_n\})\right]=\prod_n{P_n(\hat{e}_n)},
\label{EQ2-Theo}
\end{equation}
where $P_n(\hat{e}_n)$ are the single-site probabilities
\begin{equation}
P_n(\hat{e}_n)=\frac{\exp\left[\beta\textbf{h}_n\cdot\hat{e}_n\right]}{Z_{0,n}},
\label{EQ3-Theo}
\end{equation}
and $Z_0$ is the corresponding partition function,
\begin{equation}
\begin{aligned}
Z_0 & =\prod_n{Z_{0,n}} \\
& =\prod_n{\int{\text{d}\hat{e}_n\exp\left[\beta\textbf{h}_n\cdot\hat{e}_n\right]}} 
=\prod_n{4\pi\frac{\sinh\beta h_n}{\beta h_n}}.
\end{aligned}
\label{EQ4-Theo}
\end{equation}
These equations can then be used to carry out the average over $\{\hat{e}_n\}$ of quantities of interest. For example, the thermally averaged orientation at site $n$ is
\begin{equation}
\left\{\textbf{m}_n=\int{\text{d}\hat{e}_nP_n(\hat{e}_n)}\hat{e}_n=\left(\frac{-1}{\beta h_n}+\coth\beta h_n\right)\hat{h}_n\right\},
\label{EQ5-Theo}
\end{equation}
which by definition corresponds to the single-site magnetic order parameter. In addition, the entropy can be calculated as
\begin{equation}
\begin{aligned}
S_\text{mag} & =\sum_n{S_n}(\beta h_n) \\
& =-k_\text{B}\sum_n\int{\text{d}\hat{e}_n P_n(\hat{e}_n)\log P_n(\hat{e}_n)} \\
& =k_\text{B}\sum_n\left[1+\log\left(4\pi\frac{\sinh\beta h_n}{\beta h_n}\right)-\beta h_n\coth \beta h_n\right].
\end{aligned}
\label{EQ6-Theo}
\end{equation}

Invoking the Peierls-Feynman inequality~\cite{Feynman1955} we obtain an upper-bound $G_1$ of the exact Gibbs free energy as
\begin{equation}
G_1=\langle\tilde{\Omega}(\{\hat{e}_n\})\rangle_0-TS_\text{mag}.
\label{EQ7-Theo}
\end{equation}
where $\langle\cdots\rangle_0$ stands for the average with respect to $\{P_n(\hat{e}_n)\}$. Since the average of the exact Hamiltonian is prescribed by the single-site probabilities, $\langle\tilde{\Omega}(\{\hat{e}_n\})\rangle_0$ naturally depends on the local order parameters $\{\textbf{m}_n\}$. Several full DFT-based disordered local moment 
theory calculations~\cite{Mendive-Tapia2017,Staunton2014,Zemen2017} have shown that this average is well described by
\begin{equation}
\begin{aligned}
\langle\tilde{\Omega}(\{\hat{e}_n\})\rangle_0=
-\sum_{n,n'}\mathcal{J}_{nn'}
(\textbf{m}_{n}\cdot\textbf{m}_{n'}) \\
-\sum_{n,n'}\sum_{n'',n'''}\mathcal{K}_{nn'n''n'''}
(\textbf{m}_{n}\cdot\textbf{m}_{n'})
(\textbf{m}_{n''}\cdot\textbf{m}_{n'''}),
\end{aligned}
\label{EQ8-Theo}
\end{equation}
where $\mathcal{J}_{nn'}$ and $\mathcal{K}_{nn'n''n'''}$ are constants describing pair-wise and the lowest possible order of multi-site interactions connecting the magnetic order parameters, respectively. Note that Eq.\ (\ref{EQ8-Theo}) preserves the symmetry $\{\textbf{m}_n\}\rightarrow-\{\textbf{m}_n\}$ since these quantities change sign under time reversal. We have found that this expression qualitatively and quantitatively fits our ab-initio DFT-DLM calculations which describe incommensurate magnetism in the heavy rare earth elements~\cite{Mendive-Tapia2017}, the antiferromagnetic-ferromagnetic metamagnetic phase transition in FeRh~\cite{Staunton2014}, as well as the first-order paramagnetic-triangular phase transition and the effect of biaxial strain on the magnetic frustration in Mn$_3$GaN~\cite{Zemen2017}. The key point is that the quartic coefficients $\mathcal{K}_{nn'n''n'''}$ describe multi-site interactions or, equivalently, the effect of magnetic ordering growth on the underlying itinerant electronic structure and how the consequent changes on the spin-polarized electronic structure can affect and qualitatively alter the interactions between the local moments. We use the insights from these calculations to set the generic form of our model Gibbs free energy for Mn$_3$NiN.

For the Mn$_{3}A$N antiperovskites, where there are three Mn atoms inside one unit cell for the triangular phase, the associated magnetic order parameters have equal sizes and form angles of 120 degrees in the (111) lattice plane of the $\Gamma^{4g}$ and $\Gamma^{5g}$ structure (or small additional out-of-plane deviations in the $\Gamma^{4g}$ structure). By naming them as $M_1$, $M_2$, and $M_3$, we can exploit the symmetry and define the following constants
\begin{equation}
\begin{aligned}
\frac{1}{3}A_2(\Delta\omega)=\sum_{nn'}\mathcal{J}_{nn'}(\Delta\omega)\cos(\theta_{nn'}),
\end{aligned}
\label{EQ9a-Theo}
\end{equation}
\begin{equation}
\begin{aligned}
& \frac{1}{3}A_4(\Delta\omega)= \\ 
& \sum_{nn'}\sum_{n''n'''}\mathcal{K}_{nn'n''n'''}
(\Delta\omega)\cos(\theta_{nn'})\cos(\theta_{n''n'''}),
\label{EQ9b-Theo}
\end{aligned}
\end{equation}
where $\theta_{nn'}$ is the angle between the magnetic order parameters at sites $n$ and $n'$ . Note that they have an explicit dependence on the pressure induced relative volume change $\Delta\omega = \frac{\mathrm{d}V}{V}$, where $V$ is the volume. In principal the $\mathcal{J}_{nn'}$ and $\mathcal{K}_{nn'n''n'''}$ constants could be obtained from ab-initio calculations for Mn$_{3}$NiN. These calculations are ongoing but we are developing the theory to include the effects of the faster itinerant spin fluctuation effects associated with the Ni sites alongside the slower local moment fluctuations on the Mn sites. In the present work we instead identify their effect in the more compact form of $A_2$ and $A_4$ from our experimental data in Mn$_{3}$NiN and Mn$_{3}$GaN and verify their existence.

The final steps in the establishment of our Gibbs Free energy model take into account the simplest possible magnetovolume coupling, \textit{i.e.} linear in the volume change and quadratic in the magnetic order parameters. This follows by setting $A_2=a_2+c_\text{mv}\Delta\omega$ and neglecting the volume dependence of the quartic coefficients, that is $A_4=a_4$. We finally write the Gibbs free energy $G$ as
\begin{equation}
\begin{aligned}
G = & -a_{2}(M^{2}_{1} + M^{2}_{2} + M^{2}_{3}) -a_{4}(M^{4}_{1} + M^{4}_{2} + M^{4}_{3}) \\
 & - c_{\mathrm{mv}}\Delta\omega(M^{2}_{1} + M^{2}_{2} + M^{2}_{3}) \\
 & + \frac{1}{2}\Omega\gamma\Delta\omega^{2} + \Omega\Delta\omega p - TS_{\mathrm{tot}}.
\end{aligned}
\label{EQ11-Theo}
\end{equation}
where the last term is the total entropy composed by the entropy contributions from the three sub-lattices, $S_{\mathrm{tot}} = S_{1} + S_{2} + S_{3}$, given by Eq.\ (\ref{EQ6-Theo}).
Note that we have also added a simple elastic term proportional to the inverse of the compressibility, $\gamma$ (with a value around $\sim 130$\,GPa for the Mn-based antiperovskites \cite{Zemen2017a,Takenaka2014}), and the effect of an external hydrostatic pressure, $p$. 
By minimizing Eq.\ (\ref{EQ11-Theo}) with respect to the secondary order parameter $\Delta\omega$ we obtain
\begin{equation}
\Delta\omega=\frac{1}{\gamma\Omega}\left[c_{mv}(M_1^2+M_2^2+M_3^2)-\Omega p\right],
\label{EQ11b-Theo}
\end{equation}
which substituted into Eq.\ (\ref{EQ11-Theo}) gives
\begin{equation}
\begin{aligned}
G = & -\left[a_{2} - \frac{c_{\mathrm{mv}}}{\gamma} p\right] (M^{2}_{1} + M^{2}_{2} + M^{2}_{3}) \\ 
 & -a_{4}(M^{4}_{1} + M^{4}_{2} + M^{4}_{3}) - \frac{c^{2}_{\mathrm{mv}}}{2\Omega\gamma} (M^{2}_{1} + M^{2}_{2} + M^{2}_{3})^2 \\
 & - \frac{\Omega p^{2}}{2\gamma} - TS_{\mathrm{tot}}
\end{aligned}
\label{EQ12-Theo}
\end{equation}
By making use of Eqs.\ (\ref{EQ5-Theo}) and (\ref{EQ6-Theo}), the free energy expressed above can be minimized with respect to the magnetic order parameters for given values of $T$ and $p$ and consequently the theoretical quantities of \TN, $|\frac{\mathrm{d}T_{\mathrm{N}}}{\mathrm{d}p}|$ and $\Delta\omega_{T_{\mathrm{N}}}$ can be suitably calculated as functions of the model parameters $a_2$, $a_4$, and $c_{mv}$.

To conclude this section, we derive the condition for which the transition from the paramagnetic sate ($M_1=M_2=M_3=0$) to the triangular state ($M_1=M_2=M_3\neq 0$) changes from second order to first order. As shown by Bean and Rodbell~\cite{Bean1962}, this condition follows from requiring that the fourth-order coefficient in the order parameters of $G$ is negative. To find this we firstly expand Eq.\ (\ref{EQ6-Theo}) in terms of $m_n$ using Eq.\ (\ref{EQ5-Theo}),
\begin{equation}
S_n=k_\text{B}\left(\log 4\pi-\frac{3}{2}m_n^2-\frac{9}{20}m_n^4-\dots\right).
\label{EQ13-Theo}
\end{equation}
Now from Eqs.\ (\ref{EQ12-Theo}) and (\ref{EQ13-Theo}) it directly follows that the condition is
\begin{equation}
a_4=\frac{3}{10}a_2-\frac{3c_{mv}^2}{2\Omega\gamma}.
\label{EQ14-Theo}
\end{equation}

\begin{figure}
\includegraphics[width=0.45\textwidth]{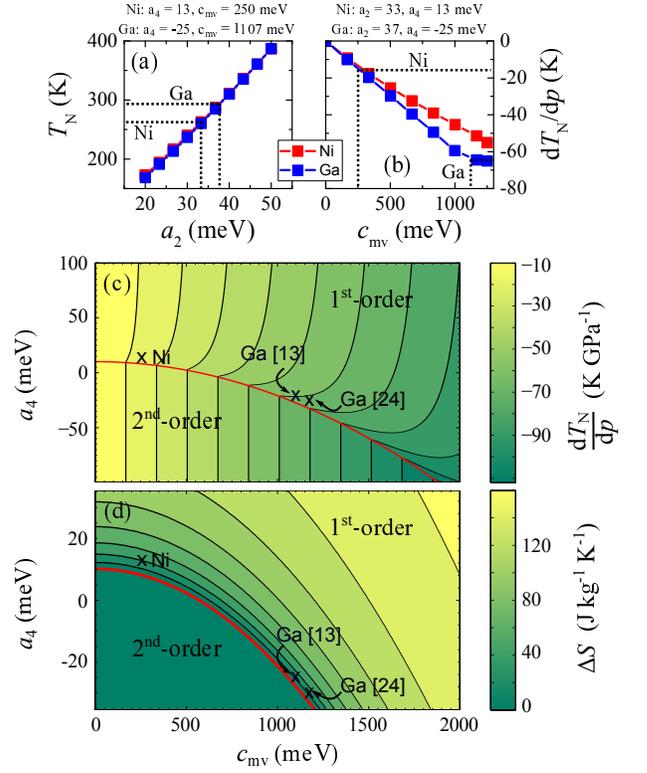}%
\caption{The theoretical quantities (a) $T_{\mathrm{N}}$ and (b) $|\frac{\mathrm{d}T_{\mathrm{N}}}{\mathrm{d}p}|$ as a function of $a_{2}$ and $c_{\mathrm{mv}}$, respectively, calculated using our model described in the text. Dotted lines indicate the experimentally measured values for $A =$ Ni and Ga. The parameters used are shown above each plot.
(c-d) The dependence of (c) $|\frac{\mathrm{d}T_{\mathrm{N}}}{\mathrm{d}p}|$ and (d) $\Delta S$ as a function of $a_{4}$ and $c_{\mathrm{mv}}$ with $a_{2} = 36$\,meV for both nitride systems.
The plot is effectively insensitive to the $a_{2}$ value for the pertinent range ($33 < a_{2} < 37$\,meV).  The red line indicates the critical curve separating $1^{\mathrm{st}}$- and $2^{\mathrm{nd}}$-order behaviour.}
\label{theory-fig}
\end{figure}

\begin{table*}
\caption{Experimental values of \TN, $\Delta\omega_{T_{\mathrm{N}}}$ and  $\frac{\mathrm{d}T_{\mathrm{N}}}{\mathrm{d}p}$ for Mn$_{3}$NiN and Mn$_{3}$GaN along with the calculated values of $c_{\mathrm{mv}}$, $a_{2}$, $a_{4}$ and $\Delta S$ as described in the text. The latter, $\Delta S$, is calculated from the Clausius-Clapeyron equation but in excellent agreement with that determined from calorimetry. The maximum magnetic entropy available for a $S = \frac{5}{2}$ Mn ion, commonly given by $\Delta S = k_{\mathrm{B}}\mathrm{ln}(2S+1)$, in Mn$_{3}$NiN and Mn$_{3}$GaN is 188.2 and 179.8\,J\,kg$^{-1}$\,K$^{-1}$, respectively. As such the Mn$_{3}A$N family have the potential to realise much larger entropy changes.}
\begin{ruledtabular}
\begin{tabular}{c | ccccccc}
 & $T_{\mathrm{N}}$ (K) & $\Delta\omega_{T_{\mathrm{N}}}$ (\%) & $\frac{\mathrm{d}T_{\mathrm{N}}}{\mathrm{d}p}$ (K\,GPa$^{-1}$) & $c_{\mathrm{mv}}$ (meV) & $a_{2}$ (meV) & $a_{4}$ (meV) & $\Delta S$ (J\,K$^{-1}$\,kg$^{-1}$) \\ \hline
 Mn$_{3}$NiN & 262 & 0.4 & 13.5 & 250 & 33 & 13 & 47 \\
 Mn$_{3}$GaN & 290 \cite{Matsunami2014} & 1.0 \cite{Matsunami2014} & 65 \cite{Matsunami2014} & 1107 & 37 & -25 & 22
\end{tabular}
\end{ruledtabular}
\label{table}
\end{table*}

\section{Determining the Multi-Site Interaction Strength}

By minimizing $G$ of Eq. \ref{EQ12-Theo} with respect to the magnetic order parameters for given values of $T$ and $p$, the theoretical quantities of $T_{\mathrm{N}}$, $|\frac{\mathrm{d}T_{\mathrm{N}}}{\mathrm{d}p}|$, and $\Delta \omega_{T_{\mathrm{N}}}$ can be calculated as a function of the model parameters $a_{2}$, $a_{4}$ and $c_{\mathrm{mv}}$. 
For instance, we show in Fig. \ref{theory-fig}(a) and \ref{theory-fig}(b) how $T_{\mathrm{N}}$ and $|\frac{\mathrm{d}T_{\mathrm{N}}}{\mathrm{d}p}|$ depend on $a_{2}$ and $c_{\mathrm{mv}}$, respectively, whilst dependencies of $a_{4}$ on (i) $c_{\mathrm{mv}}$ and $|\frac{\mathrm{d}T_{\mathrm{N}}}{\mathrm{d}p}|$ and (ii) $c_{\mathrm{mv}}$ and $\Delta S$ are shown in Fig. \ref{theory-fig}(c) and Fig. \ref{theory-fig}(d), respectively. Using this model, combined with our experimental data on Mn$_{3}$NiN and literature data on Mn$_{3}$GaN \cite{Matsunami2014}, the constants $a_{2}$, $a_{4}$ and $c_{\mathrm{mv}}$ are determined by an iterative fitting process for both systems. The final parameters are shown in Table \ref{table}.

Mn$_{3}$GaN has a significantly larger $c_{\mathrm{mv}}$ of 1107\,meV compared to the small value of 250\,meV for Mn$_{3}$NiN, consistent with the larger volume change at the transition. We propose that the five times larger ratio $|\frac{\mathrm{d}T_{\mathrm{N}}}{\mathrm{d}p}|$ observed experimentally in Mn$_{3}$GaN mainly originates from the larger magnetovolume coupling. Turning now to the $a_{2}$ and $a_{4}$ terms, we observe that $a_{2}$ scales mainly with \TN\ while positive $a_{4}$ contributes to the 1$^{\mathrm{st}}$-order character of the transition (see Equation \ref{EQ12-Theo}), in agreement with other models based on expanding the free energy in terms of the magnetic order parameter \cite{Bean1962,Mendive-Tapia2015}. Therefore, whilst $\Delta\omega_{T_{\mathrm{N}}}$ at the transition is smaller in Mn$_{3}$NiN compared to Mn$_{3}$GaN, the entropy change can be at least as large due to the smaller but positive $a_{4}$ term preserving a substantial $\Delta\omega_{T_{\mathrm{N}}}$ relative to the low $|\frac{\mathrm{d}T_{\mathrm{N}}}{\mathrm{d}p}|$ from the small magnetovolume coupling. This is exemplified in Fig. \ref{theory-fig}(c), which shows the dependence of $|\frac{\mathrm{d}T_{\mathrm{N}}}{\mathrm{d}p}|$  on $a_{4}$ and $c_{\mathrm{mv}}$. It is clear from this plot that $|\frac{\mathrm{d}T_{\mathrm{N}}}{\mathrm{d}p}|$ scales with $c_{\mathrm{mv}}$, whilst increasing positive $a_{4}$ leads to more 1$^{\mathrm{st}}$-order behaviour. Hence, from this we understand that Ni is pushed towards the region displaying 1$^{\mathrm{st}}$-order behaviour, and therefore larger $\Delta S$ (see Fig. \ref{theory-fig}(d)). 
As the $a_{4}$ term originates from the multi-site interactions among the Mn local moments and the itinerant electron spin it appears consistent that the larger BCE in Mn$_{3}$NiN relative to Mn$_{3}$GaN is due to these multi-site terms providing an additional contribution that strongly favours the triangular state. This information provides a useful handle on the barocaloric properties of Mn$_{3}A$N and stimulates research into those with small magnetovolume coupling that may have been overlooked. For instance, $A =$ Co, Pd and Rh all have significantly lower magnetovolume effects than $A =$ Ni \cite{Takenaka2014}, but their magnetism may have a contribution from the multi-site exchange that enhances the BCE based on their potential to hybridise with the Mn $d$-band (as is the case for Ni) \cite{Zemen2017a}. Moreover, it will be possible to tailor quaternary compounds that offer large BCE at room temperature \cite{Takenaka2014}.

\section{Conclusion and Outlook}

In conclusion, we have measured the barocaloric properties of the geometrically frustrated antiferromagnet Mn$_{3}$NiN, a member of the Mn-based antiperovskite family Mn$_{3}A$N. This material displays large pressure-driven isothermal entropy changes and adiabatic temperature changes of $\Delta S = 35$\,J\,K$^{-1}$\,kg$^{-1}$ ($p = 0.28$\,GPa) and $\Delta T = 5.8$\,K ($p = 0.56$\,GPa) near the paramagnetic to antiferromagnetic transition. The former is larger than the closely related antiperovskite Mn$_{3}$GaN by a factor of 1.6. Considering that the magnetovolume coupling is a factor of 5 smaller in Mn$_{3}$NiN the large entropy change is somewhat unexpected, however our theoretical insights reveal that their properties are strongly linked to a combination of electronic spin effects and lattice changes. 
The interactions between the local moments associated with the magnetic Mn sites depend on the overall magnetic order, described by the multi-site interactions. Their positive value, which is set by our experimental data, acts to enhance the $1^{\mathrm{st}}$-order nature of the transition in Mn$_{3}$NiN and consequently the material retains a large volume change despite the relatively small magnetovolume coupling. This improved understanding can be used to tune the BCE in Mn$_{3}A$N materials and highlights the potential of finding enhanced properties in this broad and chemically flexible family as well as frustrated magnets in general.

\begin{acknowledgments}
We thank A.S. Wills for help with sample preparation. This work was supported by EPSRC (UK) grants (EP/P511109/1, EP/P030548/1, EP/J06750/1 and EP/M028941/1), MINECO project no. FIS2014-54734-P and ERC Starting grant no. 680032. X.M. is grateful for support from the Royal Society.
\end{acknowledgments}


%

\newpage

\begin{widetext}
\newcommand{\beginsupplement}{%
        \setcounter{table}{0}
        \renewcommand{\thetable}{S\arabic{table}}%
        \setcounter{figure}{0}
        \renewcommand{\thefigure}{S\arabic{figure}}%
     }

\beginsupplement

\section{Supplementary Information}

\begin{figure}[h]
\includegraphics[width=0.45\textwidth]{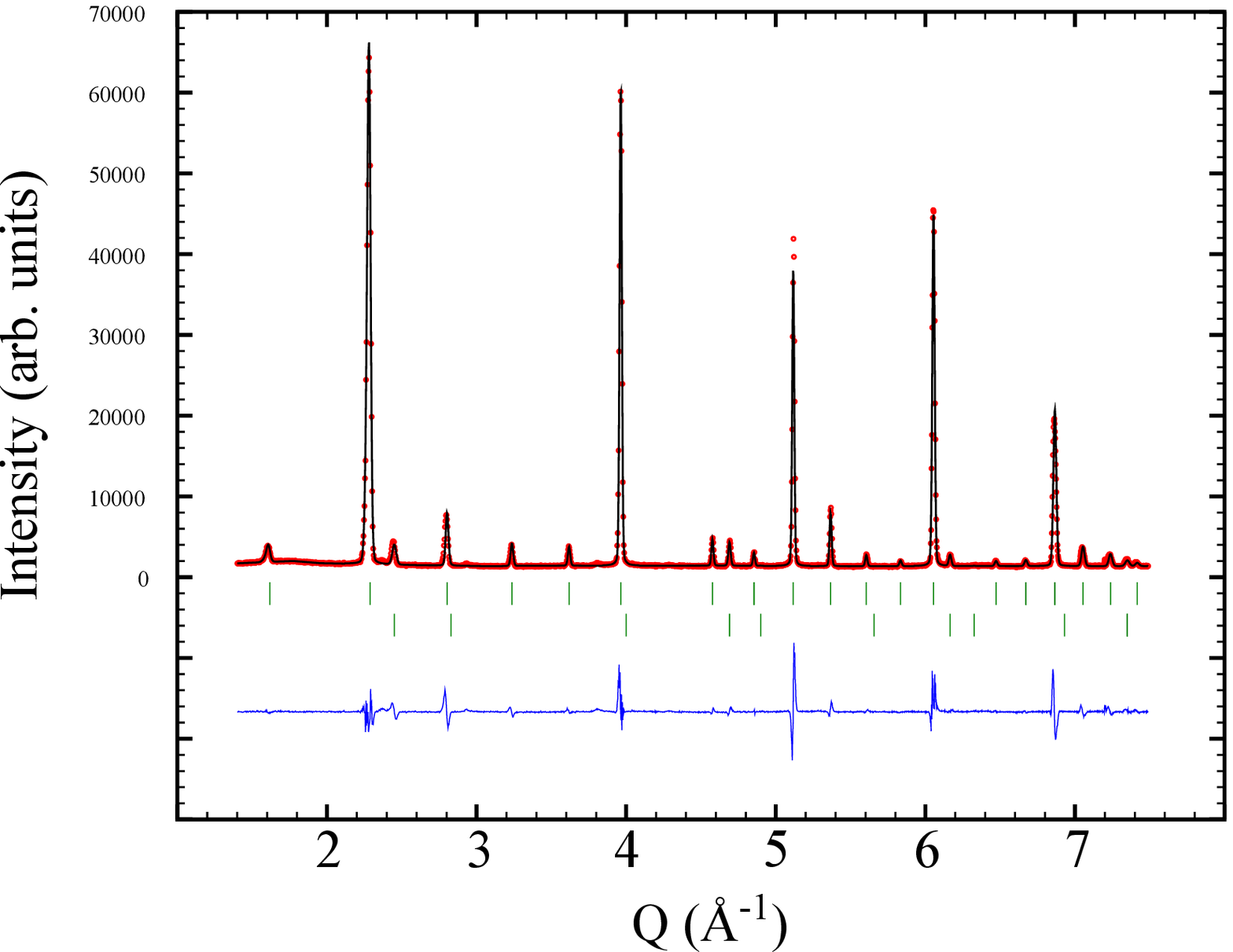}%
\caption{\label{} Rietveld refinement of neutron powder diffraction data collected on the D20 diffractometer, ILL at 285K with neutrons of $\lambda = 1.544$\,\AA. Red, black and blue lines are the measured, calculated and difference plots, respectively. The upper green tick marks correspond to the Mn$_{3}$NiN phase and the lower ones to MnO.}
\end{figure}

\subsection{Experimental details}

Magnetic measurements were performed using the vibrating sample magnetometer (VSM) option in a Quantum Design Physical Property Measurement System (PPMS-9T). For the high-pressure measurements a CuBe pressure cell was used with Daphne 7373 oil as a transmitting medium and Pb wire as an internal manometer. 

Powder neutron diffraction data was collected on the high flux diffractometer D20 at the ILL neutron source with neutrons of wavelength $\lambda = 1.544$\,\AA. The sample was loaded in a standard vanadium can which was mounted in an orange ILL cryostat. Rietveld refinement of the nuclear and magnetic structure was performed using the FullProf software package. 

Heat capacity measurements at atmospheric pressure were performed using a TA Q2000 DSC. High-pressure calorimetry was performed in a MV1-30 high-pressure cell (Unipress, Poland) adapted as a calorimeter by using peltier modules, operating up to 0.6\,GPa and from 193\,K to 393\,K. The sample was mixed with an inert perfluorinated fluid (Galden, Bioblock Scientifics) to remove air and encapsulated within Sn capsules. Two different capsules, using powder and bulk samples respectively, were measured independently. DW-Therm M90.200.02 fluid (Huber K\"{a}ltemaschinenbau GmbH) was used as a pressure transmitting medium in the pressure circuit.

\subsection{Entropy calculations}

The entropy curve $S - S_{T=210K}$ at normal pressure appearing in Fig. 1(h) has been calculated simply as the integral of the heat flow, $\frac{dQ}{dT}$, over T, that appears in Fig. 1(g):

\begin{equation}
\begin{aligned}
S(T) - S(T_{0}) = \int^{T}_{T_{0}} \frac{(\frac{\mathrm{d}Q}{\mathrm{d}T})_{p=0}}{T} dT
\end{aligned}
\end{equation}

At high pressures, our device cannot account for the heat flow outside the transition, but it is only able to detect the latent heat. Then we have to make use of the heat flow that does not correspond to the transition (\emph{i.e.} the $C_{p}$ of each individual phase). Therefore, more precisely, the entropy curves $S - S_{T=210\mathrm{K}}$ at different pressures (those appearing in Fig. 3(c) from which the BCE of Fig. 3(d) and 3(e) have been calculated) are constructed following the method used in Stern-Taulats \emph{et al.}, Appl. Phys. Lett. \textbf{107}, 152409 (2015) \cite{Stern-Taulats2015b}: Above and below the transition, the entropy is calculated using $\frac{\mathrm{d}Q}{\mathrm{d}T} = C_{p}$ in the expression above. In the transition temperature range, we take into account the contribution of the pressure-dependent heat flow after baseline subtraction (\emph{i.e.} heat flow corresponding to the latent heat of the $1^{\mathrm{st}}$-order transition, $\frac{\mathrm{d}Q}{\mathrm{d}T}/T$, shown in Fig. 3(a)), but also we include a contribution of the $C_{p}$. This has to be done this way because we do not have a strictly isothermal transition but it takes place in a finite temperature interval, and therefore the heating/cooling of the sample is costly also because of the $C_{p}$ of each phase. Therefore, in general in the transition range this $C_{p}$ has contributions from both phases such that each contribution has to be weighted according the corresponding transformed phase fraction. In particular, we take values of the $C_{p}$ outside the transition extrapolated within the temperature range (\emph{i.e.} avoiding the peak values) of the transition. Of course, this temperature range changes with pressure. Mathematically:

\begin{equation}
S(T,p) - S(T_{0},p) = 
\left\{
\renewcommand{\arraystretch}{1.2}
\setlength{\arraycolsep}{0pt}%
\begin{array}{ r>{{}}l @{\quad} c @{\quad} r>{{}}l @{\quad} l }
&\int^{T}_{T_{0}} \frac{C^{\mathrm{II}}(T^{\prime})}{T^{\prime}}dT^{\prime} 
&T \leq T_{1} \\
&S(T_{1},p) + \int^{T}_{T_{1}} \frac{1}{T^{\prime}} \left(xC^{\mathrm{II}}_{\mathrm{p}} + (1-x)C^{\mathrm{I}}_{\mathrm{p}} + |\frac{dQ(T^{\prime},p)}{dT}|\right)dT^{\prime} 
&T_{1} < T \leq T_{2} \\
&S(T_{2},p) + \int^{T}_{T_{2}} \frac{C^{\mathrm{I}(T^{\prime})}}{T^{\prime}}dT^{\prime}
&T_{2} < T
\end{array}
\right.
\end{equation}

\noindent
where the transition at a given pressure occurs between $T_{1}$ and $T_{2}$. Here, $x$ is the fraction of the system transformed to phase II and can be extracted from the calorimetric peaks at high pressure (Fig. 3(a)). Notice that the values of $C_{p}$ that are used within the transition range are those extrapolated from the baseline of the $C_{p}$ in Fig. 1(g), \emph{i.e.} avoiding the values belonging to the peak, as these are taken into account through $\frac{dQ}{dT}$ obtained from high-pressure calorimetry.

%

\end{widetext}

\end{document}